\documentclass[conference]{IEEEtran}
\IEEEoverridecommandlockouts
\usepackage{cite}
\usepackage{amsmath,amssymb,amsfonts}
\usepackage{algorithmic}
\usepackage{graphicx}
\usepackage{textcomp}
\usepackage{graphicx}
\usepackage{multirow}
\usepackage{booktabs}
\usepackage{multirow}
\usepackage{xcolor}
\def\BibTeX{{\rm B\kern-.05em{\sc i\kern-.025em b}\kern-.08em
    T\kern-.1667em\lower.7ex\hbox{E}\kern-.125emX}}
\begin{document}

\title{Memory-Based Malware Detection under Limited Data Conditions: A Comparative Evaluation of TabPFN and Ensemble Models \\
}


\author{\IEEEauthorblockN{Valentin Leroy
\IEEEauthorrefmark{1}, Shuvalaxmi Dass
\IEEEauthorrefmark{2}, Sharif Ullah
\IEEEauthorrefmark{3}\\
    \IEEEauthorblockA{\IEEEauthorrefmark{1}CESI Engineering School, Arras, France}
    \IEEEauthorblockA{\IEEEauthorrefmark{2}University of Louisiana at Lafayette, Lafayette, LA, USA}
    \IEEEauthorblockA{\IEEEauthorrefmark{3}University of Central Arkansas, Conway, AR, USA}
    Email: \texttt{valentin.leroy1@viacesi.fr}, \texttt{shuvalaxmi.dass@louisiana.edu}, \texttt{mullah@uca.edu}
}
\thanks{This is the author-accepted manuscript of a paper accepted for publication in IEEE ICNC. The final published version will appear in IEEE Xplore.}
}

\maketitle

\begin{abstract}
Artificial intelligence and machine learning have significantly advanced malware research by enabling automated threat detection and behavior analysis. However, the availability of exploitable data is limited, due to the absence of large datasets with real-world data. Despite the progress of AI in cybersecurity, malware analysis still suffers from this data scarcity, which limits model generalization. In order to tackle this difficulty, this work investigates TabPFN, a learning-free model designed for low-data regimes. We evaluate its performance against established baselines—Random Forest, LightGBM, and XGBoost—across multiple class configurations. Our experimental results indicate that TabPFN surpasses all other models in low-data regimes, with a 2\% to 6\% improvement observed across multiple performance metrics. However, this increase in performance has an impact on its computation time in a particular case. These findings highlight both the promise and the practical limitations of integrating TabPFN into cybersecurity workflows.
\end{abstract}


\section{Introduction}
Malware detection is still a major problem in the field of cybersecurity since contemporary malicious software is using more and more techniques like obfuscation, encryption, and polymorphism to avoid detection. Traditional methods, such as static and dynamic analysis, have worked well in controlled environments, but they frequently don't adapt well to real-world situations where the variety of malware types and their corresponding behaviors are always changing \cite{memory2}.

Machine learning (ML) has emerged as a promising solution for detecting and classifying malicious activities, leveraging models such as Random Forest (RF), Support Vector Machines (SVM), Gradient Boosting (XGBoost), and LightGBM, which have shown strong performance in malware and intrusion detection tasks \cite{memory2}. Although ML-based malware research has advanced rapidly, many existing studies \cite{Gupta2022, Firat2022} depend heavily on large quantities of labeled data or handcrafted feature engineering for model development. This dependence significantly restricts the real-world applicability of such approaches, particularly in situations characterized by data scarcity or extreme class imbalance. In cybersecurity, obtaining sufficient, high-quality real-world malicious samples poses a uniquely complex challenge \cite{Paul2025}. Barriers such as privacy regulations and the rapid evolution of cyberattacks further hinder the creation of comprehensive, representative labeled datasets essential for effective ML-based cybersecurity.

To address these limitations, researchers have increasingly explored tabular learning models for cybersecurity applications, where structured data (such as system logs or memory features) plays a key role. However, most prior works have primarily focused on binary classification (malware vs. benign) \cite{memory}, providing little insight into the ability of models to discriminate between multiple malware families—a crucial requirement for fine-grained threat analysis \cite{memory, memory2}.

Driven by these difficulties, this study intends to explore the capabilities of TabPFN, a newly developed probabilistic transformer specifically for tabular data, which is engineered to excel with small datasets and does not necessitate hyperparameter adjustment \cite{b1}.
We compare its performance against established ensemble models such as Random Forest, LightGBM, and XGBoost using the CIC-MalMem-2022 dataset under multiple class granularities (3-, 10-, and 15-class scenarios). Throughout this work, it should be noted that the \textbf{ensemble models} refer to the collection of state-of-the-art ML models included in our evaluation.

Our objective is to evaluate whether TabPFN can maintain strong predictive performance and generalization capabilities as the number of malware families increases, while also analyzing its computational efficiency in data-limited environments.
The main contributions of this work can be summarized as follows:

\begin{itemize}
\item We conduct a comprehensive evaluation of TabPFN compared to other state-of-the-art machine learning algorithms, namely Random Forest, XGBoost, and LightGBM. This comparison aims to verify whether TabPFN can deliver superior performance across different classification granularities.
\item We evaluate the impact of class granularity and training size on both the predictive performance and computational efficiency of TabPFN compared to traditional tree-based models.
\end{itemize}

The remainder of this paper is organized as follows:
Section II presents the Related Work and positions our study within the existing literature. Section III describes the Hardware Setup used in our experiments. Section IV details the Methodology, including data preparation, sampling, and validation strategies. Section V reports and discusses the Results, and Section VI concludes the paper with remarks and directions for future work.

\section{Related Work}
The field of malware detection has been extensively studied. However, modern types of malware are constantly changing by using methods like obfuscation and polymorphism. These changes make traditional detection techniques less effective. Recent studies have looked into machine learning methods, but many still depend on manually created features or small datasets, limiting their real-world applicability.

Carrier \textit{et al.}\cite{memory} proposed a memory-based detection framework enhancing \textit{VolMemLyzer} with 26 new features to identify obfuscated threats. Their work show almost a 99\% accuracy on a small 3-class dataset (Trojan, Spyware, Ransomware).  
However, its performance remains constrained by the limited dataset size, low class diversity, and reliance on handcrafted features, which restrict the model’s ability to generalize to broader and more complex scenarios. Dener \textit{et al.}\cite{memory2} applied memory analysis in a big-data setting using the \textit{CIC-MalMem-2022} dataset and evaluated models such as Random Forest, Gradient Boosted Trees, Logistic Regression, and LSTMs. And achieving up to 99.97\% accuracy for binary malware-versus-benign classification.  
While these results are impressive, their focus remains exclusively on binary detection, without addressing the more complex challenge of multi-family classification or scalability across growing class sets.

Other studies have emphasized the effectiveness of tree-based ensemble models for tabular malware datasets (Random Forest, XGBoost, Light GBM), motivating their use as strong baselines.  
More recently, Villafranca \textit{et al.} \cite{b1} demonstrated that \textit{TabPFN}, a probabilistic transformer for tabular learning, can outperform traditional models in Industrial IoT intrusion detection, particularly in low-data regimes.  
However, their evaluation was limited to fewer than ten classes and not tailored to memory-based malware datasets. Despite these advances, the issue of data scarcity remains a persistent challenge in the cybersecurity domain. Most available malware datasets are either small, synthetic, or lack real-world diversity, making it difficult to train generalizable models. For example, MalDroid-2020 mainly contains synthetic data and therefore fails to represent the diversity of real-world Android threats, while EMBER relies solely on static features of Windows executables, limiting its capacity to model runtime behaviors \cite{EMBER}, \cite{Shafiq}. Building on this line of research, our work focuses on improving malware detection under limited-data conditions, where collecting large and representative datasets is often infeasible in real-world cybersecurity scenarios. While most prior works have primarily addressed binary classification, we extend the task to multi-class settings involving up to 15 malware families to further evaluate model robustness. To achieve this, we compare modern tabular models (TabPFN, LightGBM, XGBoost, and Random Forest) to assess their scalability and performance when trained on small, realistic subsets of data. Previous studies \cite{b1}, \cite{c2}, \cite{c3} have demonstrated the strong performance of tree-based ensemble methods for tabular malware analysis. This motivates our evaluation of TabPFN, which is specifically designed for small-data regimes, to determine whether it can deliver tangible improvements in accuracy and inference efficiency as data availability decreases and class complexity increases.

\section{Methodology}

This section details the methodological framework followed in this study. It consists of multiple steps, including the dataset preparation, sampling strategy, model configuration, and evaluation process. 

\subsection{Dataset Preparation and Class Balance}
\begin{figure}[htbp]
    \centering
    \includegraphics[width=0.43\textwidth]{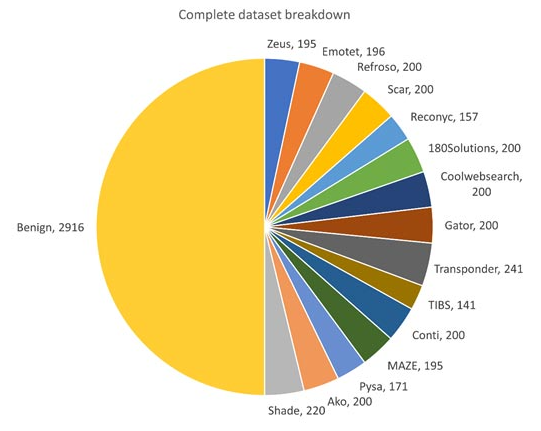} 
    \caption{Complete dataset breakdown.}
    \label{fig}
\end{figure}

The dataset employed in this work is malware memory analysis (CIC-MalMem-2022) \cite{b4}. It contains a total of 58,596 samples, evenly split between benign and malware samples (Figure \ref{fig}).
However, while the benign class is well represented, the malware portion is highly imbalanced across 15 distinct families. Given that our focus is on fine-grained malware family classification, we excluded benign samples to prevent them from artificially skewing the class distribution and inflating the metric values.

The dataset contains no missing values; however, there are some issues with the family names in the target column. To address this, we created a new column, `Family', derived from the `Category' column, to accurately assign the family/group name for each malware sample. In our work, we aim to adapt TabPFN and other machine learning models to handle datasets with varying class sizes. Specifically, we will focus on three scenarios involving 3, 10, and 15 classes.

\begin{itemize}
\item For the 3-class setting, we utilized the same dataset in which families were pre-grouped into three broad categories: Spyware, Malware, and Trojan.
\item In the 10-class setting, we have experimented with multiple combinations of malware categories, each containing 5 classes. In order to represent the best outcome, we decided to exclude all families belonging to the Spyware group.
\item For the 15-class setting, we will consider all available families.
\end{itemize}

\subsection{Sampling Bias}
The detection of malware in practical applications frequently encounters the issue of restricted and imbalanced data accessibility. In order to replicate this realistic limitation and to analyze the performance of models in such situations, we intentionally conduct sampling on the initial dataset. To avoid sampling bias, we used stratified sampling when selecting our subsets, ensuring that all data set splits preserve the same class distribution. This guarantees that the proportion of each malware family remains consistent across training, testing, and validation sets. For the whole subset, we will use a fraction of the whole dataset. Only 3000 samples will be taken from the original one. The distribution can be seen in the Table \ref{tab:train_test_sizes}.

And to be precise, the testing set corresponds to the remaining portion of the dataset — that is, the total (2,500 samples) minus the percentage used for training.




\subsection{Validation \& Split size}
For hyperparameter tuning, we will utilize 500 balanced samples and perform five-fold cross-validation to determine the optimal hyperparameter settings. This strategy minimizes overfitting and ensures efficient use of limited data, resulting in more stable and generalizable outcomes. Beyond comparing the performance of various machine learning algorithms, the experiment also investigates how their behavior and accuracy evolve as the training data size increases. The objective is to evaluate the scalability and data sensitivity of these algorithms, with particular attention to their performance when trained on smaller datasets.

\vspace{-4mm}
\begin{table}[h!]
\caption{Training and testing split sizes}
\vspace{-2mm}
\label{tab:train_test_sizes}
\centering
\begin{tabular}{|c|c|c|c|}
\hline
\textbf{Training size(\%)} & \textbf{Training} & \textbf{Testing} & \textbf{Validation} \\ \hline
10\% & 250  & 2250 & 500 \\ \hline
20\% & 500  & 2000 & 500 \\ \hline
30\% & 750  & 1750 & 500 \\ \hline
50\% & 1250 & 1250 & 500 \\ \hline
\end{tabular}
\vspace{-3mm}
\end{table}

\subsection{Validation Metrics }

To evaluate our models, we will rely on the standard performance metrics:  \\
\[
accuracy = \frac{TP + TN}{TP + TN + FN + FP}
\]
\[
precision = \frac{TP}{TP + FP}
\]
\[
recall = 2.\frac{TP}{TP + FN}
\]
\[
F1 score = \frac{Precision. Recall}{Precision + Recall}
\]

TP: True positives,
TN: True negatives,
FP: False positives,
FN: False negatives. 

Furthermore, to measure TabPFN’s ability to handle dataset with different levels of multiclasses, we will also evaluate the models in terms of both training time and inference time (test).

\subsection{Fine Tuning}
In this work, we opted to keep TabPFN in its original, unmodified configuration, meaning no hyperparameter tuning was performed on the model. We made this decision because TabPFN is specifically designed to be used without fine-tuning, and we wanted to preserve its intended usage without altering its parameters. In contrast, baseline models (Random Forest, LightGBM, and XGBoost) need tuning on the hyperparameter to have better performance. To ensure a fair and rigorous comparison, we conducted a Random Search over few key hyperparameters for each baseline model. We rely on the same hyperparameter settings as described in \cite{b1}. This strategy enables the exploration of a broad hyperparameter space and helps identify high-performing configurations, allowing us to evaluate each model under comparable experimental conditions. The results showed a slight advantage for Grid Search, but at the cost of significantly longer execution time, while the performance gain was marginal (around 0.8\%). Therefore, we chose Random Search as it offers a much faster search time with nearly identical performances (Table \ref{tab1}). 


\vspace{-2mm}
\begin{table}[htbp]
\caption{Comparison of Randomized Search, Grid Search, and untuned configuration on 500 samples with Random Forest.}
\vspace{-3mm}
\begin{center}
\begin{tabular}{|c|c|c|c|c|}
\hline
\textbf{Model}&\multicolumn{4}{|c|}{\textbf{Metrics score}} \\
\cline{2-5} 
\textbf{} & \textbf{\textit{Accuracy}}& \textbf{\textit{Precision}}& \textbf{\textit{Recall}}& \textbf{\textit{F1 score}} \\
\hline\hline
RF Grid Search     & 0.3207 & 0.3241 & 0.3183 & 0.3120 \\
\hline
RF Randomize Search& 0.3147 & 0.3139 & 0.3119 & 0.3054 \\
\hline
RF                 & 0.2708 & 0.2668 & 0.2679 & 0.2611 \\
\hline
\end{tabular}
\label{tab1}
\end{center}
\vspace{-3mm}
\end{table}

In addition, we are going to focus on the optimization of the F1 score in priority. And to evaluate the impact of the hyperparameters, we focused on the 10-class configuration for the results. The results presented in the Table \ref{tab1} do not aim to contribute to the core findings of this paper, but rather serve to illustrate the differences between the methods.

\section{Evaluation \& Results}
This section presents the hardware set up,  experimental results and analysis of all evaluated models. It covers hyperparameter tuning outcomes, computational time measurements, and comparative performance across multiple class configurations.

\subsection{Hardware set‑up}
To ensure that our work can be reproduced by others, all experiments will be done using Google Colab, a cloud-based platform that provides a pre-configured environment to execute Python code. The GPU provides is a NVIDIA Tesla T4 GPU, which is available for free in Colab. With the TabPFN model, it is highly recommended to use a GPU rather than a CPU \cite{a1}. For that reason, to guarantee computational efficiency, all models are executed on the GPU.

\subsection{Experimental Setup}
For the training and testing process, we use a Stratified Shuffle Split \cite{split}. This method preserves the class distribution in each partition while repeating the procedure multiple times, allowing us to report the average performance metrics over a predefined number of splits. For all train–test configurations (10–90\%, 20–80\%, 30–70\%, and 50–50\%), we perform 5 shuffles in every multiclass situations. This choice ensures consistency across experiments while providing a reliable average of the deferent metrics. 

\subsection{Hyperparameter Tuning}
On average, hyperparameter tuning resulted in an improvement of approximately 2–4\% across all evaluation metrics, with slight variations depending on the data subset considered.

\vspace{-2mm}
\begin{table}[htbp]
\caption{Performance of the models on 500 samples ($^{*}$ denotes finetuned models)}
\label{tab2}
\vspace{-3mm}
\centering
\begin{tabular}{c c c c c}
\hline
\textbf{Models} & \multicolumn{4}{c}{\textbf{Metric Score}} \\
\cline{2-5}
 & \textbf{Accuracy} & \textbf{Precision} & \textbf{Recall} & \textbf{F1-score} \\
\hline\hline
LGBM$^{*}$ & 0.3287 & 0.3296 & 0.3246 & 0.3172 \\
LGBM       & 0.3068 & 0.2953 & 0.3036 & 0.2933 \\
RF$^{*}$   & 0.3229 & 0.3278 & 0.3191 & 0.3108 \\
RF         & 0.2929 & 0.3008 & 0.2907 & 0.2839 \\
XGBoost$^{*}$ & 0.3049 & 0.3044 & 0.3013 & 0.2950 \\
XGBoost       & 0.3188 & 0.3263 & 0.3166 & 0.3109 \\
\hline
\end{tabular}
\vspace{-2mm}
\end{table}

Table \ref{tab2} shows the comparative performance of finetuned models with the models with default configurations. 
Interestingly, in our experiments, the default (untuned) XGBoost configuration achieved higher performance than the tuned versions. This outcome can be explained by several factors. First, XGBoost's default hyperparameters are designed to provide a well-balanced trade-off between bias and variance, especially on small to medium-sized datasets. In contrast, hyperparameter tuning on a limited dataset may overfit to the cross-validation folds, leading to reduced generalization on unseen data. Moreover, exploring a very large search space with randomized search can cause the optimization process to select suboptimal configurations due to insufficient coverage within the limited number of iterations. Finally, the default configuration implicitly applies moderate regularization, which helps stabilize training, while tuned configurations may reduce this regularization and thus overfit.
These factors together explain why, in some cases, the untuned XGBoost model can outperform its tuned counterpart. The final hyperparameter values obtained from the tuning process are summarized in Table \ref{tab:hyperparameters_all}.


\begin{table}[h!]
\centering
\caption{Hyperparameter Settings for Each Model}
\label{tab:hyperparameters_all}
\resizebox{0.3\textwidth}{!}{
\begin{tabular}{llc}
\toprule
\textbf{Model} & \textbf{Hyperparameter} & \textbf{Value} \\
\midrule
\multirow{5}{*}{LightGBM}
 & learning\_rate & 0.043 \\
 & max\_bin & 20 \\
 & max\_depth & 20 \\
 & num\_leaves & 30 \\
 & random\_state & 42 \\
\midrule
\multirow{5}{*}{Random Forest}
 & n\_estimators & 365 \\
 & max\_depth & 28 \\
 & max\_features & 1 \\
 & criterion & entropy \\
 & random\_state & 42 \\
\midrule
\multirow{5}{*}{XGBoost}
 & learning\_rate & 0.421 \\
 & eta & 0.363 \\
 & max\_depth & 4 \\
 & subsample & 0.828 \\
 & random\_state & 42 \\
\bottomrule
\end{tabular}
}
\vspace{-2mm}
\end{table}

\subsection{Computational Time Analysis}
When trained with different sample configurations, XGBoost emerged as the best-performing model, achieving significantly lower training and testing times compared to the other models in most cases (Table \ref{tab:times}).

\begin{table}[htbp]
\centering
\caption{Training and inference time comparison.}
\vspace{-1mm}
\label{tab:times}
\resizebox{0.4\textwidth}{!}{%
\begin{tabular}{ c l | c | c | c}
\hline
\multicolumn{2}{c|}{} & 3-Class & 10-Class & 15-Class \\
\hline
Training size & Algorithm & Time (s) & Time (s) & Time (s) \\
\hline
250 (10\%)  & RF       & 0.8           & 2.33          & 3.08 \\
     & XGBoost  & \textbf{0.34} & \textbf{0.75} & \textbf{0.99} \\
     & LightGBM & 0.39          & 1.34          & 1.46 \\
     & TabPFN   & 2.34          & 2.61          & 39.55 \\
\hline
500 (20\%)  & RF       & 0.89          & 2.08          & 2.88 \\
     & XGBoost  & \textbf{0.31} & \textbf{0.87} & 1.18 \\
     & LightGBM & 0.57          & 0.97          & \textbf{1.11} \\
     & TabPFN   & 3.0           & 2.8           & 47.7 \\
\hline
750 (30\%)  & RF       & 0.97          & 2.25          & 3 \\
     & XGBoost  & \textbf{0.31} & \textbf{0.9}  & \textbf{1.29} \\
     & LightGBM & 0.75          & 1.22          & 1.41 \\
     & TabPFN   & 3.42          & 3.24          & 56.12 \\
\hline
1250 (50\%) & RF       & 1.13          & 2.54          & 3.26 \\
     & XGBoost  & \textbf{0.33} & \textbf{0.96} & \textbf{1.35} \\
     & LightGBM & 0.74          & 1.68          & 1.88 \\
     & TabPFN   & 4.17          & 4.01          & 67.03 \\
\hline

\end{tabular}%
}
\vspace{-5mm}
\end{table}

\begin{table*}[htbp]
\centering
\caption{Performance comparison on 3-Class, 10-Class, and 15-Class classification tasks.}
\vspace{-1mm}
\label{tab:results}
\resizebox{\textwidth}{!}{%
\begin{tabular}{c l | cccc | cccc | cccc}
\hline
\multicolumn{2}{c|}{} & \multicolumn{4}{c|}{3-Class Classification} & \multicolumn{4}{c|}{10-Class Classification} & \multicolumn{4}{c}{15-Class Classification} \\
\hline
Training set size & Algorithm & Accuracy & Precision & Recall & F1 score & Accuracy & Precision & Recall & F1 score & Accuracy & Precision & Recall & F1 score \\
\hline
250  & RF       & 0.5345 & 0.5346 & 0.5345 & 0.5332 & 0.2515 & 0.248 & 0.2496 & 0.2468 & 0.2193 & 0.2236 & 0.2204 & 0.2193 \\
     & XGBoost  & 0.5273 & 0.5271 & 0.5272 & 0.526 & 0.2353 & 0.2342 & 0.2346 & 0.2329 & 0.2021 & 0.2117 & 0.2028 & 0.2044 \\
     & LightGBM & 0.5206 & 0.5202 & 0.5203 & 0.5197 & 0.2323 & 0.2303 & 0.2314 & 0.2293 & 0.202 & 0.2121 & 0.203 & 0.2047 \\
     & TabPFN   & \textbf{0.5623} & \textbf{0.564} & \textbf{0.5626} & \textbf{0.5583} & \textbf{0.2641} & \textbf{0.2765} & \textbf{0.26} & \textbf{0.252} & \textbf{0.2359} & \textbf{0.248} & \textbf{0.2344} & \textbf{0.2205} \\
\hline
500  & RF       & 0.5675 & 0.5674 & 0.5672 & 0.5664 & 0.3109 & 0.3051 & 0.3097 & 0.3054 & 0.2627 & 0.2665 & 0.2619 & 0.2613 \\
     & XGBoost  & 0.5568 & 0.556 & 0.5562 & 0.5556 & 0.2911 & 0.2868 & 0.29 & 0.2871 & 0.244 & 0.2489 & 0.2445 & 0.2446 \\
     & LightGBM & 0.5512 & 0.5511 & 0.5506 & 0.5504 & 0.2884 & 0.2859 & 0.2878 & 0.2857 & 0.245 & 0.2485 & 0.2458 & 0.2457 \\
     & TabPFN   & \textbf{0.5965} & \textbf{0.5957} & \textbf{0.5962} & \textbf{0.5943} & \textbf{0.3402} & \textbf{0.3397} & \textbf{0.337} & \textbf{0.3295} & \textbf{0.281} & \textbf{0.2884} & \textbf{0.2777} & \textbf{0.2622} \\
\hline
750  & RF       & 0.5904 & 0.5921 & 0.5902 & 0.5902 & 0.3407 & 0.3364 & 0.3401 & 0.3363 & 0.2857 & 0.288 & 0.2867 & 0.2851 \\
     & XGBoost  & 0.5747 & 0.5752 & 0.5744 & 0.5743 & 0.3171 & 0.3148 & 0.3161 & 0.3145 & 0.276 & 0.28 & 0.2777 & 0.2777 \\
     & LightGBM & 0.5603 & 0.5608 & 0.5602 & 0.5599 & 0.3176 & 0.3134 & 0.3169 & 0.3142 & 0.2693 & 0.2737 & 0.2713 & 0.2712 \\
     & TabPFN   & \textbf{0.6175} & \textbf{0.6176} & \textbf{0.6173} & \textbf{0.6161} & \textbf{0.3704} & \textbf{0.3628} & \textbf{0.3676} & \textbf{0.3585} & \textbf{0.3317} & \textbf{0.3331} & \textbf{0.3317} & \textbf{0.3185} \\
\hline
1250 & RF       & 0.6024 & 0.6024 & 0.602 & 0.6019 & 0.3848 & 0.3832 & 0.3848 & 0.3826 & 0.3358 & 0.3367 & 0.3345 & 0.3331 \\
     & XGBoost  & 0.5949 & 0.595 & 0.5946 & 0.5942 & 0.3626 & 0.3631 & 0.3623 & 0.3619 & 0.3165 & 0.3238 & 0.3161 & 0.3182 \\
     & LightGBM & 0.5776 & 0.5775 & 0.5772 & 0.5769 & 0.3555 & 0.355 & 0.3554 & 0.3541 & 0.303 & 0.3073 & 0.3025 & 0.3032 \\
     & TabPFN   & \textbf{0.6386} & \textbf{0.638} & \textbf{0.638} & \textbf{0.6373} & \textbf{0.416} & \textbf{0.4177} & \textbf{0.4145} & \textbf{0.4103} & \textbf{0.3725} & \textbf{0.3726} & \textbf{0.3702} & \textbf{0.3583} \\
\hline
\end{tabular}%
}
\vspace{-4mm}
\end{table*}

By contrast, TabPFN did not perform well in terms of computational efficiency, as its training and inference times were considerably higher across all scenarios—particularly in the 15-class classification task. In this setting, TabPFN was between 38.56s and 65.68s slower than the fastest baseline model. This overhead, however, can largely be attributed to the extended version of TabPFN, which remains under development \cite{a2}.

For the other classification tasks, TabPFN was competitive only with Random Forest, most notably in the 10-class setting, but did not consistently outperform the more optimized tree-based baselines. Apart from the 15-class classification task, the differences in training and inference time between TabPFN and the baseline models remain relatively modest. On average, the gap is around 2 seconds, which can be considered a minor overhead in practical scenarios.

\vspace{-1mm}
\subsection{Classification Results and Analysis}

In this phase, multiclass classification was performed using varying class sizes included in our malware samples. The detailed outcomes are presented in Table \ref{tab:results}. 

\subsubsection{3‑Class classifcation}
The results of the 3-class classification show that TabPFN outperforms the baseline models across all evaluation metrics. The improvement ranges from 2\% to 6\%, depending on the compared model. However, as previously mentioned, these gains come at the cost of higher computational time.

\subsubsection{10‑Class classifcation}
In the 10-class classification scenario, TabPFN remained the best-performing model, though the margins over the baselines were narrower than in the 3-class case. With 250 samples, TabPFN achieved 26.4\% accuracy compared to 25.1\% for Random Forest and 23.5\% for XGBoost. As the training size increased, the advantage became more pronounced: at 1250 samples, TabPFN reached 41.6\%, while Random Forest achieved 38.5\% and XGBoost 36.3\%. These results show that TabPFN remains competitive even as the task complexity increases, though its lead is primarily significant against Random Forest.

\subsubsection{15‑Class classifcation}
The task involving 15-class classification was the most difficult for all models, as indicated by overall lower accuracy rates. With a dataset of 250 samples, TabPFN achieved an accuracy of 23.6\%, which was a slight improvement over Random Forest at 21.9\% and XGBoost at 20.2\%. When the number of samples increased to 1250, TabPFN's accuracy rose to 37.3\%, whereas Random Forest reached 33.6\% and XGBoost attained 31.6\%. Though TabPFN maintained the highest accuracy, the improvement was less pronounced in this nuanced classification context. This situation underscores the complexity of the task and the computational trade-offs involved when utilizing TabPFN with a greater number of classes.

Overall, across all classification configurations, TabPFN consistently outperformed the baseline models in terms of accuracy and generalization. The performance gap was most noticeable in low-class and low-sample settings, confirming its strength under limited-data conditions. However, these advantages come at the expense of higher computational cost, particularly as the number of classes and data complexity increase. On the other hand, imbalanced data imposes an issue that needs a data preparation step in our approach. We intend to address these issues in the extension of this work.

\section{Conclusion \& Future Work}
In this study, TabPFN was evaluated for the purpose of classifying malware families, particularly in relation to baseline models such as Random Forest, XGBoost, and LightGBM, across various training sizes and class levels of 3, 10, and 15. The findings indicate that TabPFN, particularly under conditions of low data size scenarios, surpasses the baseline models, achieving enhancements ranging from 2\% to 6\%. However, a significant increase in computational times occurs while the class number is higher (15 classes in our case). These findings highlight the promise of TabPFN in data-scarce situations and underscore the need for further adjustments to facilitate its effective application within the domain of cybersecurity. Future work could extend this study by enhancing the performance of TabPFN for large-scale multiclass classification through exploring recent extensions, such as AutoTabPFN, to improve adaptability and efficiency.


\begin{thebibliography}{00}

\bibitem{memory2} Dener, Murat \& Orman, Abdullah. (2022). Malware Detection Using Memory Analysis Data in Big Data Environment. Applied Sciences. 12. 23. 10.3390/app12178604

\bibitem{memory} Tristan Carrier, Princy Victor, Ali Tekeoglu, Arash Habibi Lashkari,” Detecting Obfuscated Malware using Memory Feature Engineering”, The 8th International Conference on Information Systems Security and Privacy (ICISSP), 2022

\bibitem{b1} S. Ruiz‑Villafranca, J. Roldán‑Gómez, J. Manuel Castelo Gómez, J. Carrillo‑Mondéjar, J. Luis Martinez, `A TabPFN‑based intrusion detection system for the industrial internet of things' 

\bibitem{EMBER} Anderson, H. and Phil Roth. “EMBER: An Open Dataset for Training Static PE Malware Machine Learning Models.” ArXiv abs/1804.04637 (2018): n. pag.

\bibitem{Shafiq} Samaneh Mahdavifar, Andi Fitriah Abdul Kadir, Rasool Fatemi, Dima Alhadidi, Ali A. Ghorbani; Dynamic Android Malware Category Classification using Semi-Supervised Deep Learning, The 18th IEEE International Conference on Dependable, Autonomic, and Secure Computing (DASC), Aug. 17-24, 2020.

\bibitem{c2} Anjani Gupta, Dr. Karan Singh (2023) Malware Analysis on AI Technique

\bibitem{c3} Benjamin Marais, Tony Quertier, St´ephane Morucci (2022) AI-based Malware and Ransomware Detection Models

\bibitem{a1} PriorLabs GitHub repository TabPFN -
https://github.com/PriorLabs/TabPFN

\bibitem{b4} CIC-MalMem-2022 - Dataset -
https://www.unb.ca/cic/datasets/malmem-2022.html

\bibitem{split} Pedregosa, F., Varoquaux,  \& Duchesnay, É. (2011). Scikit-learn: Machine Learning in Python. Journal of Machine Learning Research, 12, 2825–2830.

\bibitem{a2} PriorLabs GitHub repository TabPFN extensions -
https://github.com/PriorLabs/tabpfn-extensions

\bibitem{Gupta2022} C. Gupta, I. Johri, K. Srinivasan, Y.C. Hu, S.M. Qaisar, and K.Y. Huang,
'A Systematic Review on Machine Learning and Deep Learning Models for Electronic Information Security in Mobile Networks,'

\bibitem{Firat2022}I. Firat et al.,'Cyber risk and cybersecurity: a systematic review of data availability,'

\bibitem{Paul2025}B. Paul,'Machine Learning for Cybersecurity Issues: A Systematic Review,' 2025.

\end{thebibliography}
\end{document}